\documentclass{article}

\usepackage{epsfig}
\usepackage{multirow}

\usepackage[nofiglist,notablist,nomarkers]{endfloat}

\def\deg{\ensuremath{^{\circ}}}
\renewcommand{\u}[1]{\ensuremath{\mathrm{#1}}}

\begin{document}

\title{The Cosmic Electron Background in Low Energy IACTs. 
Effect of the Geomagnetic Field}
\author{J. Cortina
\footnote{
Corresponding author.\hspace{20cm}
E E-mail address: cortina@mppmu.mpg.de,
Tel. +49 89 323 54 445, Fax +49 89 322 67 04}, 
J. C. Gonz\'alez\\
{\small Max Planck Institut f\"ur Physik, F\"ohringer Ring
6, D-80805, Munich, Germany,}} 

\maketitle
\thispagestyle{empty}
%
\begin{abstract}
A new generation of low threshold Imaging Atmospheric
Cherenkov Telescopes (IACTs) may reach $\gamma$-ray energies about
10 GeV with high sensitivities and very large collection
areas. At these low energies cosmic electrons 
significantly contribute to the telescope background and are
in principle indistinguishable from $\gamma$-rays. In this
paper we estimate the electron background expected for two configurations
of the low energy IACT MAGIC. We discuss in particular the
reduction of the background caused by the geomagnetic field
at different locations on the Earth's surface.
\end{abstract}

\vspace{0.5cm}
\begin{center}
{\small
{\it PACS}: 95.55.Ka\\
\vspace{2mm}
{\it Keywords}: $\gamma$-rays, atmospheric Cherenkov detectors, \\
sensitivity, electron background, geomagnetic, GeV-TeV energies.
}
\end{center}

\vspace{2cm}
\begin{center}
Accepted for publication in Astroparticle Physics.
\end{center}

\newpage

%
%
\section{Introduction}
%
%
A number of low energy threshold Imaging Atmospheric
Cherenkov Telescopes are under design or construction 
and will cover the $\gamma$-ray
region from 10 GeV to 10 TeV. Ground-based instruments
\cite{iacts99}
provide high sensitivities and huge collections areas 
of the order of 10$^5$ m$^2$, i.e. orders of magnitude higher
than those achievable by satellite experiments in the same
energy range like GLAST \cite{kniffen99}. 
\par
In this paper we shall concentrate on the IACT MAGIC 
\cite{bradbury95,barrio98} since this
project aims at the lowest
energy threshold. The MAGIC collaboration is currently
building a 17 meter diameter telescope equipped with a $\sim$
600 PMT pixel camera (MAGIC phase I) which will
reach an energy threshold of $\sim$ 30 GeV \cite{barrio98}
(defined as the peak in the raw differential rate, i.e., 
before any $\gamma$ selection cuts).
In its second phase the camera will be equipped with high
quantum efficiency hybrid PMTs (HPD) that will
lower the energy threshold down to $\sim$ 10 GeV. A third
phase may use a camera with Avalanche PhotoDiodes (APDs) 
of even higher efficiency, but we will not
consider this for our studies.
\par
IACTs have to deal with a background of electron- and hadron-initiated
atmospheric showers. At very low energies below 50 GeV
hadrons rarely generate enough Cherenkov light to trigger
the telescope. According to \cite{barrio98,gonzalez00} its
contribution to the background is negligible below 30
GeV. In turn cosmic electrons will give rise to showers which are
practically indistinguishable from $\gamma$-induced ones
and thus can not be eliminated using traditional
$\gamma$-hadron discrimination methods. They are isotropic
and can be suppressed by means of an improved angular
resolution when the telescope searches for point sources, 
but they constitute an irreducible background in
extended source searches and diffuse gamma flux estimations.
\par
The purpose of this paper is to estimate the fluxes and
detection rates of cosmic electrons expected in the new
generation IACT MAGIC in its phases I and II. In particular
we shall carefully consider the effect of the geomagnetic
field on the electron background observed at ground level.
%
%
\section{Geomagnetic Field and Rigidity Cutoffs}
\label{geofield}
%
%
The Geomagnetic Field (GF) is a combination of several
magnetic fields produced by different sources. More than
90\% of the field is generated inside the Earth and is known
as ``main field'' or ``internal field''. The main field is
usually described as a dipole (as in the original work of
St\"ormer \cite{stormer30}) or more precisely as a high
order harmonic expansion whose coefficients are based on
experimental observations and updated every few years, the
so-called International Geomagnetic Reference Field (IGRF).
\par
The solar wind interacts with the Earth's field and its
effect is described by an external field. The currents
induced in the magnetosphere and ionosphere may create
magnetic fields as large of 10\% of the main field and are
variable on a much shorter time scale. Magnetohydrodynamic
models are called for in order to describe the whole GF,
such as Tsyganenko model \cite{tsyganenko89} or the
Integrated Space Weather Prediction Model (ISM)
\cite{erickson91,lambour94}.
The GF bends the cosmic ray trajectories preventing low 
rigidity particles from reaching the Earth's surface. 
(The rigitidy of a particle is defined as {\sl pc/ze}, where
{\sl c} is the speed of light, {\sl p} is the momentum 
and {\sl ze} is the charge of the particle).
The minimum allowed rigidity is know as rigidity cutoff. In the
case of a dipole GF the rigidity cutoffs can be calculated
analytically \cite{stormer30}. In
this paper we will draw mainly on an expression for the
rigidity cutoff ($R_c$) given by Cooke\cite{cooke83} which is based
on a dipole approximated to the IGRF1980:
\begin{equation}
\label{eq_cooke}
R_c = \frac{59.4 ~ \cos^4\lambda}
{r^2 [1+(1-\cos^3\lambda ~ \sin\theta ~ \sin\phi)^{1/2}]^2} ~ \u{GV}
\end{equation}
where $\lambda$ is the magnetic latitude, $\theta$ is the zenith
angle, $\phi$ is the azimuth angle measured clockwise from
the magnetic north and $r$ is the distance from the dipole center (in
Earth radius units). Analytical procedures always underestimate
$R_c$ due to the effect of the Earth shadow
effect. A subset of
the trajectories intersect the Earth and are physically
impossible. This results in a region of rigidities some of
which are allowed and some forbidden, usually referred to as
penumbra. The penumbra can be studied by ways of numerical
procedures of backtracking \cite{shea67,flueckiger91}.
The highest forbidden energy, so-called upper rigidity
cutoff, is 5--30\% higher than $R_c$
\cite{shea65,lipari95}.
\par
$R_c$ values in the zenith direction
according to equation \ref{eq_cooke} are tabulated in table 
\ref{tab_cutoffs} for some geomagnetic latitudes.
MAGIC will actually be built very close to 
a geomagnetic latitude of 30\deg N.
Included in the table are the extreme cases of the
magnetic pole (90\deg\ magnetic latitude), 
for which all trajectories are allowed, and the
magnetic equator (0\deg\ magnetic latitude) 
for which the cutoff energy reaches a maximum.
%
%
\section{The Electron Background}
%
%
%
\subsubsection*{Electron detection rates}
In the following we will make the assumption that
electron-initiated showers behave exactly like
gamma-initiated ones. 

\begin{figure}[p]
\begin{center}
\epsfig{file=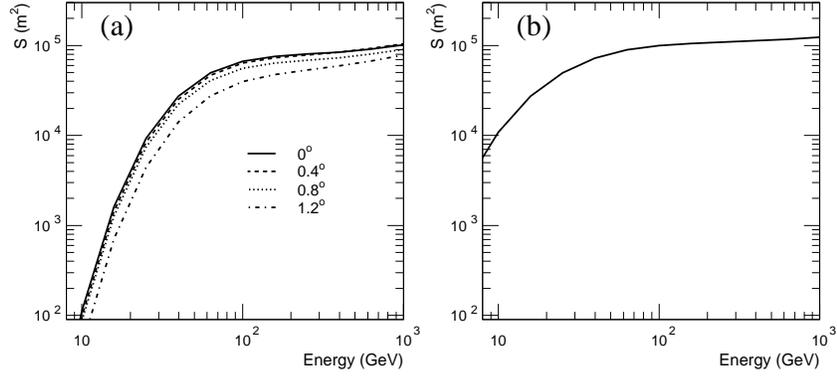,width=\textwidth}
\caption{\it 
Collection areas at zenith for
$\gamma$-initiated showers for a) MAGIC phase I 
and b) MAGIC phase II.
The solid lines represent the collection area in the
telescope pointing direction (on-axis
showers) while the broken lines correspond to
different off-axis incident angles.
The uncertainties in collection area are
always below 10\%. (From \cite{gonzalez00}).
}
\label{fig_collection_areas}
\end{center}
\end{figure}

\begin{figure}[p]
\begin{center}
\epsfig{file=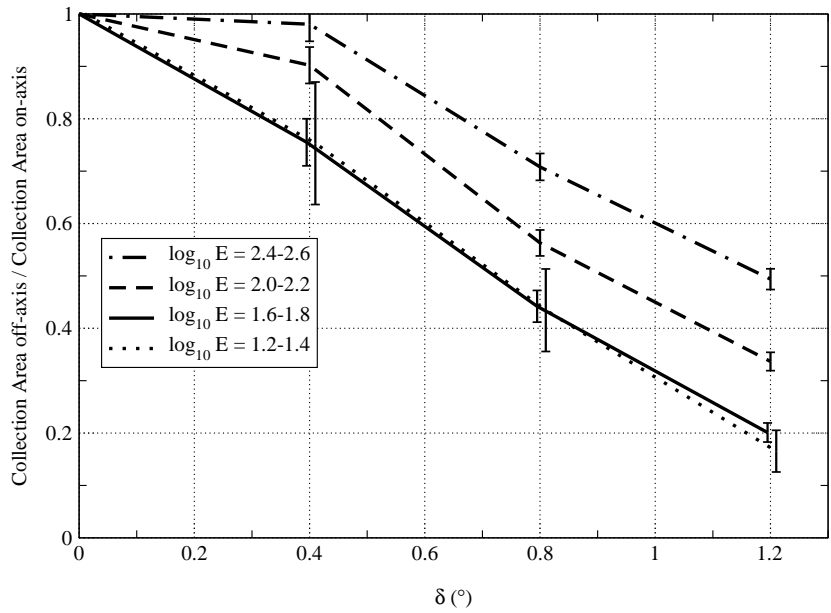,width=0.9\textwidth}
\caption{\it 
Off-axis to on-axis collection area ratio
plotted against the off-axis angle $\delta$. 
The values for four different energy regions 
(expressed in GeV) have been shown. The points 
obtained have been joined by lines to guide the eye.
}
\label{fig_area_fractions}
\end{center}
\end{figure}

Figure \ref{fig_collection_areas} shows the MAGIC
raw collection area (i.e. with no $\gamma$ selection cuts)
at zenith for gamma primaries in the case
of the PMT camera (MAGIC phase I, figure
\ref{fig_collection_areas}a) and of the HPD camera (MAGIC phase
II, figure \ref{fig_collection_areas}b). 
The collection area for $\gamma$ primaries 
was calculated using a Monte Carlo
simulation of the shower development in the atmosphere and
the response of the detector to the Cherenkov light produced
in the shower (see \cite{barrio98,gonzalez00} for details).
A trigger condition of four next-neighbour pixels 
with at least 7 photoelectrons in a trigger region of 
0.8\deg\ radius was adopted.
Electrons arrive isotropically in the camera, so 
the shower axis was generated parallel to the 
telescope pointing direction (``on-axis''), but also with 
a certain angle $\delta$ with respect to this direction
(``off-axis'') in order to characterize all possible
incident directions.
In particular $\delta$=0\deg, 0.4\deg, 0.8\deg\ and 1.2\deg\ were 
simulated. 
Figure \ref{fig_collection_areas} shows the collection area 
($S_\delta$) in MAGIC phase I for these directions.
The uncertainties in $S_\delta$ are always 
below 10\%.
Figure \ref{fig_area_fractions} shows $S_\delta$
as a function of $\delta$. $S_\delta$ has been
normalized to the on-axis collection area 
($S_{\delta=0}$) in order to show how the trigger efficiency
decreases with $\delta$.
The dependence of $S_\delta$/$S_{\delta=0}$ 
on $\delta$ has been plotted for four different energies. 
For phase II no simulation of off-axis $\gamma$
showers was available, hence we assume in our calculations 
that the fractions for phase I hold valid for phase II 
and we can use them to obtain the collection 
areas off-axis out of the collection areas on-axis.
\par
\begin{figure}[p]
\begin{center}
\epsfig{file=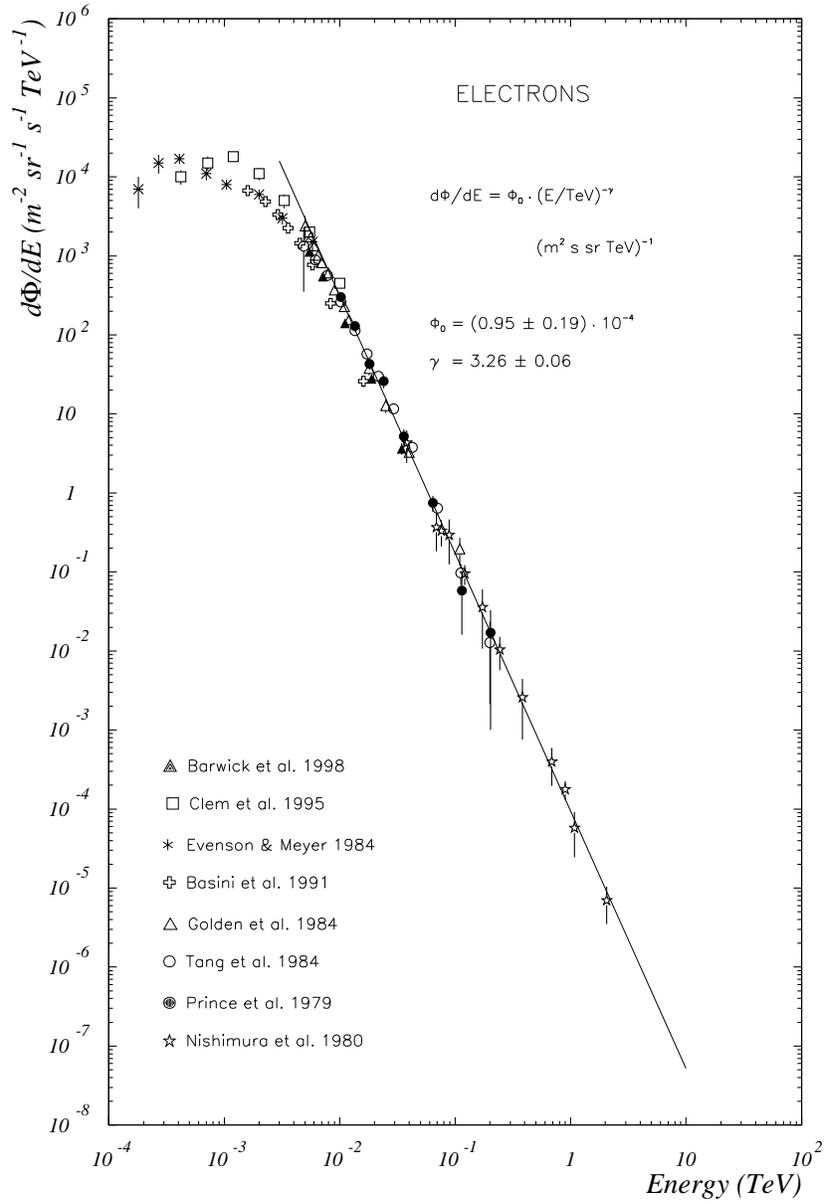,height=16cm}
\caption{\it 
Cosmic electron differential spectrum (from a
number of different experiments as summarized in \cite{wiebel98}).}
\label{fig_spectrum}
\end{center}
\end{figure}
\begin{figure}[p]
\begin{center}
\epsfig{file=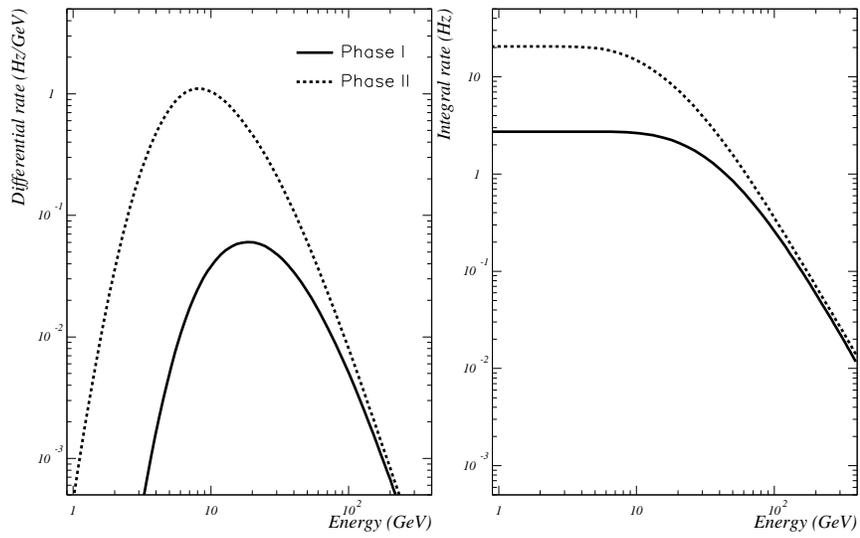,width=\textwidth}
\caption{\it 
Differential and integral raw electron rates
at the zenith position, as expected for both phases of MAGIC.}
\label{fig_rates}
\end{center}
\end{figure}

We can use the raw collection areas 
along with the electron energy spectrum displayed in figure
\ref{fig_spectrum} (taken from \cite{wiebel98}) in order to
estimate the raw rate of electron triggers in MAGIC. In doing
so we assume that electrons arrive isotropically  
with $\delta<$1.5\deg\ and that $S_\delta$ can be calculated 
from the correlations in figure \ref{fig_area_fractions}.
Figure \ref{fig_rates}
shows the electron raw differential rate and integral rate 
at zenith as a function of electron primary energy.
The uncertainties in the rates are basically determined by the
uncertainties in the primary electron spectrum which
range from $\sim$40\% at energies around 10 GeV down to $\sim$35\% 
around 1 TeV.
\par
The rate estimated for phase I (2.8$\pm$1.2 Hz) is
significantly lower than the rate quoted in
\cite{barrio98} (9 Hz). This stems from two reasons.
To begin with, we have correctly taken into account the 
decrease of collection area as a function of off-axis angle
whilst \cite{barrio98} assumed that the collection area
off-axis could be approximated by the on-axis collection area. 
Besides the authors of \cite{barrio98} did actually not quote the expected 
electron rate but rather an upper limit to this rate 
based on the upper limit to the electron
differential spectrum \cite{lorenz00}.
The electron rates obtained represent a few percent of 
the hadron rate expected for both phases I and II 
\cite{barrio98,gonzalez00}.
\par
Let us consider now the effect of the geomagnetic rigidity cutoff on
the electron rates. In table \ref{tab_final_rates}
we tabulate the electron fluxes and rates one expectes at 
different locations on
Earth as listed in table \ref{tab_cutoffs}. Also shown is the
fraction of the polar rate (rate at the magnetic pole).
The polar rate is equal to the rate that we would obtain 
neglecting the rigidity cutoff. The table displays
the rates for both phases of MAGIC pointing to the zenith position.
The errors in the rates mostly come from the uncertainties
in the primary electron flux. It must be emphasized that the
fractions of the polar rate suffer from no uncertainty
since the errors in both polar and reduced rates are correlated.
For MAGIC phase I the effect of $R_c$
is small for almost all the locations under
consideration. Only at the magnetic equator do we observe
a 25\% reduction in the rate. In contrast for MAGIC phase II
one expects a noticeable reduction in the rate, especially
for low magnetic latitudes and as high as a factor of 2 at the 
geomagnetic equator.
\par
In all previous calculations we have assumed that there is
an abrupt cut in the electron energy
spectrum at the position of the rigidity cutoff, that is,
we have ignored the penumbra. We may obtain an upper limit
to the effect of the penumbra in the electron rate by
assuming that all trajectories inside this energy region are
forbidden and that the upper rigidity cutoff is always 25\% higher
than the rigidity cutoff. In this extreme case we 
obtain further reductions in the electron rate of at most 20\%.
\subsubsection*{Telescope sensitivity}
We consider now the limits 
that the electron background pose to the detector sensitivity.
Contrary to hadron showers, electron showers behave 
exactly like $\gamma$ showers. Therefore they cannot be
rejected using separation procedures based on the 
shower image characteristics. 
For point source searches, electrons which do not come 
from the source direction may be rejected within the
angular resolution capabilities of the detector. But for diffuse or
extended sources, no rejection is possible. 
(By ``diffuse source'' we understand a source which emits 
homogenously inside our camera's field of view, while by ``extended
source'' we refer to any non-point source).
\par
The sensitivity $\Phi$ of an IACT is generally defined 
as the $\gamma$-ray flux necessary to produce a 
5$\sigma$ signal in 50 hours of observation. 
For simplicity we suppose that the $\gamma$-ray source 
has the same spectral index as the electron background.
The sensitivity for diffuse sources at 5,
10 and 50 GeV for different geomagnetic locations
can be found in table \ref{tab_sensitivity}. 
A 5\% improvement is found when going from the magnetic
pole to the magnetic equator in MAGIC phase I. 
The improvement in phase II is more considerable. The
sensitivity at the equator increases by 13\% at 10 GeV and
by 45\% at 5 GeV.
\par
In the case of point sources we can reject electron showers 
which do not come from the source direction. By using the
standard Hillas analysis methods (in particular cutting
at ALPHA parameter greater than 15\deg) we can reduce the
number of electrons by a factor of $\sim$ 6, while keeping
$\sim$ 70\% of the gammas (see page 159 of reference
\cite{barrio98}. Taking into account the orientation of 
the shower image (``head-tail''), it may be possible to reject 
50\% of the remaining electrons). 
The polar sensitivity for point source 
detection in MAGIC phase I is found to be $\Phi(>10~\u{GeV})$ = 
2.4$\cdot$10$^{-10}$ cm$^{-2}$s$^{-1}$. The
proportionality factor between this sensitivity and the
corresponding diffuse source sensitivity
can be applied to all the other sensitivities in
table \ref{tab_sensitivity} to obtain the
point-source sensitivities at 5 and 10 GeV, and, with good
approximation, also at 50 GeV.
The effect of the geomagnetic field will thus be the
same for diffuse and point sources.
\par
Most of the extended sources constitute intermediate cases
between the aforementioned point and diffuse sources.
In consequence the telescope sensitivity for a general
extended source will be determined by its actual angular 
size, but will always have a value between the 
sensitivities calculated for diffuse and point sources.
\par
The hadronic background is strongly reduced at energies 
around 10 GeV and the electron sensitivity limits which 
we have calculated are in the order of or above the limit imposed
by the hadronic background \cite{barrio98}. 
Therefore any accurate measurement 
will have to take into account the presence of the electron
background and the effect of the geomagnetic field.  
%
%
%
%
\subsubsection*{Dependence on zenith angle}
\begin{figure}[p]
\begin{center}
\epsfig{file=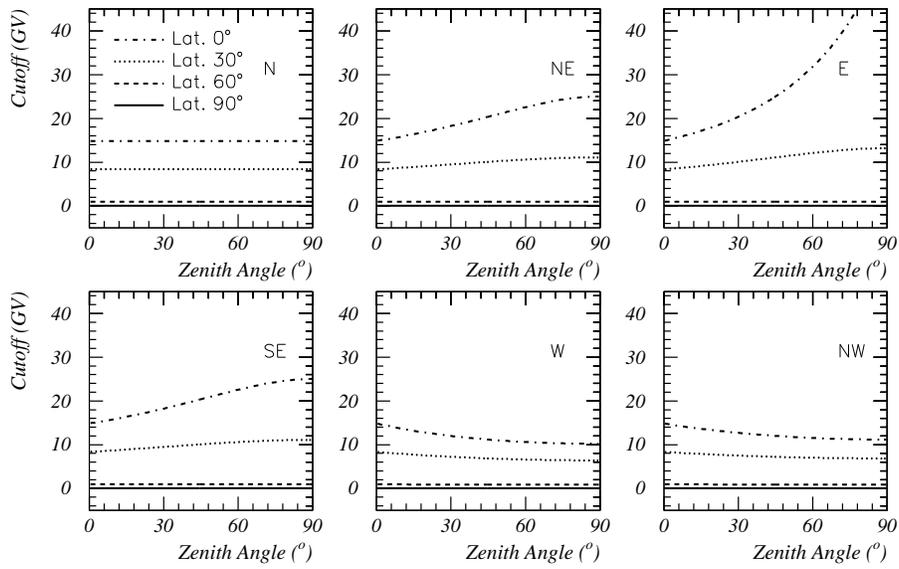,width=\textwidth}
\caption{\it 
Electron rigidity cutoff
as a function of zenith angle for six different azimuth angles,
namely in north, north-east, east, south-east, west and
north-west directions. The four curves in each figure
correspond to four magnetic latitudes: 0 \deg (magnetic
equator), 30 \deg, 60 \deg and 90 \deg
(magnetic pole). 
}
\label{fig_zenith}
\end{center}
\end{figure}

Until now we have restricted our analysis to the zenith
position of the IACTs. Let us try to predict the
geomagnetic cutoff effect at other zenith angles. The
dependence of the electron rigidity cutoff on zenith angle
for a number of azimuth angles and geomagnetic latitudes 
according to equation (\ref{eq_cooke}) is given by 
figure \ref{fig_zenith}. Whereas the cutoff does not
vary strongly with the zenith angle for geomagnetic latitudes
above 30\deg\ (at most a 50\% increase from 0 to
90\deg\ zenith angle in all azimuth directions), it
rises by as much as a factor of 3 in the geomagnetic equator.
A given IACT exhibits approximately the same 
collection area and energy threshold for all zenith angles 
below 30\deg. Hence let us apply the zenith 
collection areas shown in figure \ref{fig_collection_areas} to
all zenith angles below 30\deg. The maximal reduction in
electron rate takes place in the magnetic equator
when the telescope points to the east. The maximal
enhancement happens also in the equator when the telescope
points to the west. We have tabulated the rates expected 
for MAGIC phase II in table \ref{tab_zenith}
at geomagnetic latitudes 0\deg\
and 30\deg\ in the west and east directions. 
The maximal rate reduction is 30\%
whilst the maximal enhancement is 20\%. MAGIC phase II
sensitivity will thus improve slightly when pointing to 
the west and get slightly worse when pointing to the east. 
The effect of the zenith angle is expected to be smaller in
the case of MAGIC phase I.
\par
Conversely the IACT energy threshold increases
fast with zenith angle above 30\deg\ (see for example 
\cite{krennrich99,ibarra99,konopelko99}).
At a zenith angle of 60\deg\ the energy threshold is 
already roughly one order of magnitude higher
than the threshold close to the zenith.
Because the cosmic electron spectrum falls with primary 
energy faster than that of the cosmic ray background or than 
those of most of the predicted $\gamma$-ray sources
we can expect the electron background to be less and less
significant as we move to higher zenith angles.
In addition the effect of the geomagnetic cutoff will
decrease with zenith angle since $R_c$ does not grow
as fast as the telescope threshold energy. Electrons
triggering the telescope are thus never below $R_c$.
%
%
\section{Valididy of the Geomagnetic Model}
%
%
Equation (\ref{eq_cooke}) is based on a dipole field. Hence
we may consider to improve our calculations by applying a more
realistic main field such as the IGRF. Nevertheless
because the dipole in equation (\ref{eq_cooke}) was approximated   
to the IGRF1980 value, we do not expect our results to differ
considerably from the real values. It must be emphasized that no
analytical solution for the rigidity cutoff is possible by
using a field model more complex than a dipole field.
\par
Besides, we have always neglected in our calculations the effect of
the external geomagnetic field described in section \ref{geofield}.
As already mentioned, under normal conditions this field
contributes up to 10\% of the global field intensity.
The external field, however, is asymmetric due to the
interaction of the Earth with the solar wind, which gives
rise to a long tail extending to the direction almost opposite to
that of the Sun (actually in a direction perpendicular
to the solar wind front, but for simplicity we assume
that it extends in a direction opposite to
the Sun). This introduces an asymmetry in the electron
arrival direction, since electrons going through the tail 
are more likely to be deflected than electrons coming from
the direction of the Sun. In other words, we expect the
rigidity cutoff to grow in the direction opposite to the
Sun. For a given arrival direction this is a daily
effect. As a consequence we expect the minimum electron rate
to be reached at midnight in the zenith direction. Since
IACTs operate only at night, the electron rate will on 
the whole be further reduced by this effect.
\par
A detailed study would necessitate taking into
account the fast variations of the external field produced
during periods of increased solar activity. In general these
periods profoundly complicate the calculation of rigidity
cutoffs and background electron rates. Careful planning
of IACT observations for special studies of extended 
sources or the diffuse gamma background will be necessary 
during these periods. 
%
%
\section{Conclusions}
%
%
We have estimated the electron background expected in both
phases of the low-energy-threshold IACT MAGIC at different
geomagnetic latitudes. The raw electron rate expected for MAGIC phase I
is approximately 3 Hz in the zenith direction at the
magnetic pole and drops to 2.5 Hz at the magnetic
equator. For MAGIC phase II these rates are correspondingly
20 and 10 Hz. The reduction with decreasing magnetic
latitude reflects the fact that the electron rigidity cutoff
is minimal at high geomagnetic latitudes and maximal at the
equator. The sensitivity is also improved when going to
lower latitudes by as much as 50\% for $\Phi (>5~\u{GeV})$
at the equator. The detection rate decreases for some azimuth angles
reaching a minimum when the telescope points to the west and
increase for other angles peaking in the east direction. The
electron background is expected to be very much reduced in
high zenith angle observations.
\par
The conclusion can be drawn that any
estimate of the $\gamma$-ray flux of an extended source or
of the isotropic $\gamma$-ray flux will have to take into
account the geomagnetic location of the IACT as well as the
specific direction in which the instrument is pointing, especially
for observations in which the zenith angles are less 
than 30\deg.

%
%
\section{Acknowledgments}
%
%
The authors are deeply grateful to many members of the
MAGIC Telescope collaboration for their assistance and
interesting comments. We wish to thank especially 
R. Mirzoyan and E. Lorenz for their useful remarks and suggestions. 
One of the authors (J.C.) gratefully acknowledge
discussions with S. Orloff.

%
%

%
%
%
\begin{table}[p]
\begin{center}
\begin{tabular}{|c c|}
\hline
Magnetic latitude & Rigidity cutoff (GV)\\
\hline
\hline
Magnetic pole 90\deg & 0.\\
\hline
75\deg & 0.1\\
\hline
60\deg & 0.9\\
\hline
45\deg & 3.7\\
\hline
30\deg & 7.3\\
\hline
15\deg & 12.9\\
\hline
Magnetic equator 0\deg & 14.9\\
\hline 
\end{tabular}
\caption{\it 
Electron rigidity cutoffs in the zenith direction for
several magnetic latitudes calculated using equation
(\ref{eq_cooke}). MAGIC will actually be built very
close to 30 \deg magnetic latitude.
}
\label{tab_cutoffs}
\end{center}
\end{table}
%
%
\begin{table}[p]
\begin{center}
\begin{tabular}{|c|c|c|c|c|}
\hline
Magnetic & Rate phase I & Fraction & Rate phase II & Fraction \\
latitude & (Hz) & polar rate & (Hz) & polar rate\\
\hline
\hline
Mag. pole & 2.8$\pm$1.2 & 1.00 &  21$\pm$8 & 1.00\\
\hline
75\deg & 2.8$\pm$1.2 & 1.00 & 21$\pm$8 & 1.00\\
\hline
60\deg & 2.7$\pm$1.2 & 0.96 & 20$\pm$8 & 0.95\\
\hline
45\deg & 2.7$\pm$1.2 & 0.96 & 20$\pm$8 & 0.95\\
\hline
30\deg & 2.7$\pm$1.2 & 0.96 & 17$\pm$6 & 0.80\\
\hline
15\deg & 2.5$\pm$1.2 & 0.89 & 11$\pm$5 & 0.52\\
\hline
Mag. equator& 2.4$\pm$1.0 & 0.77 & 10$\pm$4 & 0.48\\
\hline 
\end{tabular}
\caption{\it 
Electron raw rates and fraction of the polar rate as expected for MAGIC 
phase I and II at different locations of the Earth's
surface. The polar rate is equal to the rate that we would
obtain in the absence of geomagnetic cutoff.
}
\label{tab_final_rates}
\end{center}
\end{table}
%
%
%
%
%
\begin{table}[p]
\begin{center}
\begin{tabular}{|c|c|c|c|c|}
\hline
~
 & Magn. 
 & $\Phi (>5 ~ \u{GeV})$
 & $\Phi (>10 ~ \u{GeV})$  
 & $\Phi (>50 ~ \u{GeV})$\\  
~
 & lat.
 & (cm$^{-2}$ s$^{-1}$ sr$^{-1}$) 
 & (cm$^{-2}$ s$^{-1}$ sr$^{-1}$) 
 & (cm$^{-2}$ s$^{-1}$ sr$^{-1}$) \\
\hline
\hline
\multirow{4}{13mm}{Phase I} 
        & 90\deg\ & 4.1$\cdot$10$^{-6}$ &
        1.0$\cdot$10$^{-6}$ & \multirow{4}{14mm}{4.7$\cdot$10$^{-8}$}\\
        \cline{2-4}                     
        & 30\deg\ & 4.1$\cdot$10$^{-6}$ &
        1.0$\cdot$10$^{-6}$ & \\
        \cline{2-4}                     
        & 15\deg\ & 4.1$\cdot$10$^{-6}$ &
        1.0$\cdot$10$^{-6}$ & \\
        \cline{2-4}                     
        & 0\deg\  & 3.9$\cdot$10$^{-6}$ &
        9.5$\cdot$10$^{-7}$ & \\
\hline
\hline
\multirow{4}{13mm}{Phase II} 
        & 90\deg\ & 1.6$\cdot$10$^{-6}$ & 
        4.4$\cdot$10$^{-7}$ & \multirow{4}{14mm}{3.5$\cdot$10$^{-8}$}\\
        \cline{2-4}                     
        & 30\deg\ & 1.5$\cdot$10$^{-6}$ &
        4.4$\cdot$10$^{-7}$ & \\
        \cline{2-4}                     
        & 15\deg\ & 1.2$\cdot$10$^{-6}$ &
        4.4$\cdot$10$^{-7}$ & \\
        \cline{2-4}                     
        & 0\deg\  & 1.1$\cdot$10$^{-6}$ &
        3.9$\cdot$10$^{-7}$ & \\
\hline 
\end{tabular}
\caption{\it 
Sensitivities at 5, 10 and 50 GeV for diffuse $\gamma$-ray 
sources in both phases of MAGIC. The table shows the magnetic pole 
and the magnetic latitudes for which some improvement in
sensitivity is expected at 5 or 10 GeV. 
}
\label{tab_sensitivity}
\end{center}
\end{table}
%
%
%
%
\begin{table}[p]
\begin{center}
\begin{tabular}{|c|c|c|c|}
\hline
Location & Azimuth & 0\deg\ ZA & 30\deg\ ZA \\
\hline
\hline
\multirow{2}{14mm}{Lat. 30\deg} & East & 
\multirow{2}{10mm}{17$\pm$6} & 14$\pm$5\\
\cline{2-2}
\cline{4-4}
 & West & & 18$\pm$6\\
\hline 
\hline 
\multirow{2}{14mm}{Lat. 0\deg} & East & 
\multirow{2}{10mm}{10$\pm$4} & 7$\pm$3\\
\cline{2-2}
\cline{4-4}
 & West & & 12$\pm$5\\
\hline 
\end{tabular}
\caption{\it 
Electron raw rates expected in the east and the west directions for
locations at geomagnetic latitudes 0 \deg and 30
\deg for MAGIC phase II.
}
\label{tab_zenith}
\end{center}
\end{table}

%
%
%
%
%
%
%
%
%
%
%
%
%
%
%
%
%
%
%
\mbox{}


\end{document}